\documentclass[pra,floatfix,amsmath,twocolumn,showpacs,superscriptaddress]{revtex4}
\usepackage[ansinew]{inputenc}
\usepackage{times}
\usepackage{verbatim}
\usepackage{epsfig}
\usepackage{pst-all}
\usepackage{color}
\usepackage{dcolumn}
\usepackage{amsmath}
\usepackage{amsthm}
\usepackage{bm}
\usepackage{layout}
\usepackage{float}
\usepackage{txfonts}
\usepackage{amsfonts}
\usepackage{amssymb}%
\usepackage{subfigure}
\usepackage[ansinew]{inputenc}
\usepackage{amsmath}
\usepackage{amsthm}
\usepackage{float}
\usepackage{amsfonts}
\usepackage{amssymb}

\newtheorem{lemma}{Lemma}
\newtheorem{thm}[lemma]{Theorem}
\newtheorem{prop}{Proposition}

\begin{document}
\title{General monogamy property of global quantum discord and the application}

\author{Si-Yuan Liu }
\affiliation{Institute of Modern Physics, Northwest University, Xian
710069, P. R. China }
\affiliation{Beijing National Laboratory for Condensed Matter Physics,
Institute of Physics, Chinese Academy of Sciences, Beijing 100190, P. R. China}

\author{Yu-Ran Zhang}
\affiliation{Beijing National Laboratory for Condensed Matter Physics,
Institute of Physics, Chinese Academy of Sciences, Beijing 100190, P. R. China}

\author{Li-Ming Zhao }
\affiliation{Beijing National Laboratory for Condensed Matter Physics,
Institute of Physics, Chinese Academy of Sciences, Beijing 100190, P. R. China}

\author{Wen-Li Yang }
\email{wlyang@nwu.edu.cn }
\affiliation{Institute of Modern Physics, Northwest University, Xian
710069, P. R. China }

\author{Heng Fan}
\email{hfan@iphy.ac.cn}
\affiliation{Beijing National Laboratory for Condensed Matter Physics,
Institute of Physics, Chinese Academy of Sciences, Beijing 100190, P. R. China}
\affiliation{Collaborative Innovation Center of Quantum Matter, Beijing, P. R. China}

\date{\today}

\begin{abstract}
We provide a family of general monogamy inequalities for global quantum discord (GQD),
which can be considered as an extension of the usual discord monogamy inequality. It can be shown that
those inequalities are satisfied under the similar condition for the holding of usual monogamy relation.
We find that there is an
intrinsic connection among them.
Furthermore, we present a different type of monogamy inequality and prove that it holds under the condition that
the bipartite GQDs do not increase when tracing out some subsystems. We also study the
residual GQD based on the second type of monogamy inequality.
As applications of those quantities, we investigate the GQDs and residual GQD in
characterizing the quantum phase transition in the transverse field Ising model.
%We also define two kinds of
%monogamy deficit of GQD and give the necessary and sufficient conditions for the two
%monogamy inequality hold.
\end{abstract}


\maketitle

\section{introduction}
Quantum correlations, such as entanglement and quantum discord,
are considered as valuable resources for quantum information tasks \cite{key-1,key-2,key-3,key-5,pp1,pp2,pp3}.
They also play a key role in condensed matter physics,
see, for example, Refs. \cite{p1,p2,p3,p4,p5}.
For a bipartite case, entanglement and quantum discord have been widely
accepted as two fundamental tools to quantify quantum correlations.
In general, they are quite different from each other. The research
on quantum correlation measures was initially based on the entanglement-separability
paradigm and entanglement was considered as the unique quantum correlation
that can be used to obtain a quantum speed-up.

However, it has been recently
shown that there exist quantum computational models such as the deterministic
quantum computation with one qubit (DCQ1) protocol which contains
no entanglement but demonstrates a quantum advantage \cite{key-6,key-7,key-8}.
In this sense, entanglement
does not seem to capture all the features of quantum correlations. Therefore,
many other measures of quantum correlations have been proposed in recent years.
Quantum discord \cite{key-9,key-10}
is a widely accepted one among them. The quantum discord plays an important
role in the research of quantum correlations due to its potential applications in a
number of quantum processes, such as quantum critical phenomena
\cite{key-11,key-12,key-13,key-14}, quantum evolution under decoherence
\cite{key-15,key-16} and, as we just mentioned, the DCQ1 protocol \cite{key-17}. To quantify the
multipartite quantum correlations, generalizations of bipartite quantum discord
to multipartite states have been considered in various aspects
\cite{key-18,key-19,key-20,key-21}. It is worth noting that in Ref.~\cite{key-18},
a measure of multipartite quantum correlations named as
global quantum discord (GQD) is proposed, which can be seen as a symmetric generalization of
bipartite quantum discord to multipartite cases. As a well-defined multipartite quantum
correlation, the GQD is always non-negative and symmetric with respect to subsystem
exchange. Moreover, its applications have been illustrated by the Werner-GHZ state
and in the Ashkin-Teller model \cite{key-18}.

Now, the problem of the distribution of GQD throughout a
multipartite system arises when we use GQD as a resource for quantum information processing.
Then, the monogamy property which characterizes the restriction for sharing a resource or a quantity
is helpful to provide significant information for this issue
and deserves systematic investigation.
In general, the limits on the shareability of quantum correlations are described by
monogamy inequalities \cite{key-22}. Although the quantum correlations, such as
entanglement and quantum discord, do not always obey the monogamy relations
\cite{key-23,key-24}, the monogamy property can hold for GQD \cite{key-25} for a
wider situations. This fact shows that GQD, as a multipartite quantum correlation,
has some unique advantages.

To investigate the distribution of GQD,
we provide a family of monogamy inequalities which can be taken as an extension
of the standard monogamy inequality. We can prove these new monogamy inequalities
and show an intrinsic connection between them and the standard monogamy inequality.
On this basis, we define the corresponding monogamy deficits of these inequalities and
derive an important identity for the loss of correlation. This identity brings us the
relationship between GQD of a multipartite system and GQD of its arbitrary subsystems.
In addition, we present another trade-off inequality which is also upper bounded by
GQD of a multipartite system. It reflects how GQD is distributed in the multipartite
quantum system from a different aspect. This trade-off inequality can also be regarded as a
generalized monogamy inequality, and we call it the second class of monogamy inequality
in this paper.
We also study the residual GQD in accordance with the second monogamy inequality.
Finally, we apply the GQD, the nearest-neighbor bipartite GQDs and the residual
GQD to the transverse field Ising model as the criteria to characterize the quantum phase
transitions. This shows the importance of those quantities in physical models.
We hope that our results can stimulate more researches on the connection between quantum correlations
and the quantum phase transitions.

This paper is organized as follows. In Sec.~\ref{II}, we briefly recall the definition and
properties of GQD. In Sec.~\ref{III}, we define a family of monogamy inequalities of GQD
which can be considered as an extension of the standard monogamy inequality. We prove that
they hold under the similar condition as that of the standard monogamy inequality.
An intrinsic connection between them is also presented. In Sec.~\ref{IV}, %using the definition of monogamy deficit,
%we derive an important identity of GQD. This identity tells us an interesting
we study the relationship between GQD of an $N$-partite system and that of its subsystems
by presenting another important decomposition of the loss of correlation. In Sec.~\ref{V}, we define
the second class of monogamy inequality and demonstrate that it holds under the condition
that the bipartite GQDs do not increase when subsystems are discarded. In Sec.~\ref{VI}, we
investigate the residual GQD of two typical states which relates to the second monogamy
inequality. In Sec.~\ref{VII}, we show the sum of all the nearest
neighbor bipartite GQDs and residual GQD of the transverse field Ising model
can characterize the quantum phase transition.
In the last section, we summarize our results. All proofs of the theorems in the main text are
presented in APPENDIX.

\section{global quantum discord and its properties\label{II}}
We briefly review the definition and properties of GQD proposed in Ref.~\cite{key-18}.
The definition of global quantum discord is a generalization of bipartite symmetric
quantum discord. Consider a $N$-partite system $A_{1}$, $A_{2}$, ... , $A_{N}$ (each of
them is of finite dimension), and GQD of state $\rho_{A_{1}A_{2\cdots}A_{N}}$ is defined
as follows:
\begin{eqnarray}
D\left(A_{1}:\cdots:A_{N}\right)\equiv{\min}_{\Phi}\left[I\left(\rho_{A_{1}\cdots A_{N}}\right)-I\left(\Phi\left(\rho_{A_{1}\cdots A_{N}}\right)\right)\right],
\label{1}
\end{eqnarray}
where $\Phi\left(\rho_{A_{1}\cdots A_{N}}\right)=\sum_{k}\Pi_{k}\rho_{A_{1}\cdots A_{N}}\Pi_{k}$
with $\{\Pi_{k}=\Pi_{A_{1}}^{j_{1}}\otimes\cdots\otimes\Pi_{A_{N}}^{j_{N}}\}$
representing a set of local measurements and $k$ denoting the index string
$\left(j_{1}\cdots j_{N}\right)$. In Eq.~(\ref{1}), the multipartite
mutual information $I\left(\rho_{A_{1}\cdots A_{N}}\right)$ and $I\left(\Phi\left(\rho_{A_{1}\cdots A_{N}}\right)\right)$
are given by
\begin{eqnarray}
I\left(\rho_{A_{1}\cdots A_{N}}\right)
&=&\sum_{k=1}^{N}S\left(\rho_{A_{k}}\right)-S\left(\rho_{A_{1}\cdots A_{N}}\right),\\
I\left(\Phi\left(\rho_{A_{1}\cdots A_{N}}\right)\right)
&=&\sum_{k=1}^{N}S\left(\Phi\left(\rho_{A_{k}}\right)\right)-S\left(\Phi\left(\rho_{A_{1}\cdots A_{N}}\right)\right),
\label{3}
\end{eqnarray}
where $\Phi\left(\rho_{A_{k}}\right)=\sum_{k'}\Pi_{A_{k}}^{k'}\rho_{A_{k}}\Pi_{A_{k}}^{k'}$.

GQD is a useful multipartite quantum correlation which has many advantages:
it is symmetric with respect to subsystem exchange and non-negative for arbitrary
states. GQD has been proved to be useful in the characterization of quantum phase
transitions \cite{key-18,key-26}. Moreover, GQD can play an important role in
quantum communication since it has a useful operational interpretation. In the
absence of GQD, the quantum state simply describes a classical probability
multi-distribution so that it allows for local broadcasting of correlations \cite{key-27}.

We will apply some of the properties of GQD to
prove our proposition. The main properties of GQD can be listed as follows:
$\left(a\right)$
Given a non-selective measurement $\Phi\left(\rho_{A_{1}\cdots A_{N}}\right)$,
one obtains the loss of correlation that
\begin{eqnarray}
D_{\Phi}\left(A_{1}:\cdots:A_{N}\right)
&\equiv&I\left(\rho_{A_{1}\cdots A_{N}}\right)-I\left(\Phi\left(\rho_{A_{1}\cdots A_{N}}\right)\right)\nonumber\\
&=&\sum_{k=1}^{N-1}D_{\Phi}\left(A_{1}\cdots A_{k}:A_{k+1}\right).
\label{4}
\end{eqnarray}
Therefore, GQD is the minimum of the loss of correlation $D(A_{1}:\cdots:A_{N})=\min_{\Phi}D_{\Phi}(A_{1}:\cdots:A_{N})$.
$\left(b\right)$
For an arbitrary $N$-partite system $A_{1}$, $A_{2}$,
... , $A_{N}$ , GQD obeys the following monogamy relation
\begin{eqnarray}
D\left(A_{1}:\cdots:A_{N}\right)\geq\sum_{k=1}^{N-1}D\left(A_{1}:A_{k+1}\right),
\label{5}
\end{eqnarray}
provided that the bipartite GQDs $D\left(A_{1}\cdots A_{k}:A_{k+1}\right)$ do
not increase if subsystems are discarded, that is to say, $D\left(A_{1}\cdots A_{k}:A_{k+1}\right)\geq D\left(A_{1}:A_{k+1}\right)$.

\section{general monogamy relations for global quantum discord\label{III}}
According to the previous literature, we know that the monogamy relation
$D\left(A_{1}:\cdots:A_{N}\right)\geq\sum_{k=1}^{N-1}D\left(A_{1}:A_{k+1}\right)$
holds for all quantum states whose bipartite GQDs do not increase under discard
of subsystems \cite{key-25}. Since we have this monogamy relation, an interesting
question is whether there is a more general monogamy relation holding for GQD which
includes Eq.~(\ref{5}) as a special case. In this section, we will consider this
question and provide a family of general monogamy inequalities for GQD which can
be taken as an extension of the standard monogamy inequality. It can be shown that
they can hold and have an intrinsic connection between each other. Furthermore, we derive
an important identity of GQD and show its physical significance.
%We first introduce a
%general form of these monogamy inequalities, than we prove it under the similar
%conditions for the common monogamy relation.

\begin{thm}
For an arbitrary $N$-partite system A$_{1}$, A$_{2}$, ... , A$_{N}$ and
$1<m_{1}<m_{2}<\cdots<m_{n}<N$ with $n$ an arbitrary nonnegative integer, GQD
obeys a family of general monogamy inequalities which have the following form:
\begin{eqnarray}
D(A_{1}:\cdots:A_{N})&\geq& D(A_{1}:\cdots:A_{m_{1}})\nonumber\\
+D(A_{1}:A_{m_{1}+1}:\cdots:A_{m_{2}})
&+&\cdots+ D(A_{1}:A_{m_{n}+1}:\cdots:A_{N})\ \ \ \
\label{thm1}
\end{eqnarray}
provided that the bipartite GQDs do not increase under the discarding of subsystems.
When each item of the right hand side contains only two parties, we can obtain the
standard monogamy relation (\ref{5}) as a special case.
\end{thm}

On the basis of the above results, we can introduce the corresponding monogamy
deficit which is defined as follows: the monogamy deficit of the general monogamy
inequalities
\begin{eqnarray}
\triangle D_{G}&\equiv& D(A_{1}:\cdots:A_{N})-[D(A_{1}:\cdots:A_{m_{1}})\nonumber\\
&+&D(A_{1}:A_{m_{1}+1}:\cdots:A_{m_{2}})+D(A_{1}:A_{m_{2}+1}:\cdots:A_{m_{3}})\nonumber\\
&+&\cdots +D(A_{1}:A_{m_{n}+1}:\cdots:A_{N})],
\label{def1}
\end{eqnarray}
and the monogamy deficit of the standard monogamy inequality
\begin{eqnarray}
\triangle D_{S}\equiv D(A_{1}:\cdots:A_{N})-\sum_{k=1}^{N-1}D\left(A_{1}:A_{k+1}\right).
\label{def2}
\end{eqnarray}
Using the above definition, it is easy to prove that
\begin{eqnarray}
\triangle D_{G}\leq\triangle D_{S}.
\label{11}
\end{eqnarray}
 Since every term of $\triangle D_{G}$ obeys the standard monogamy relation, we have
\begin{eqnarray}
\begin{array}{c}
D(A_{1}:\cdots:A_{m_{1}})\geq\sum_{k=1}^{m_{1}-1}D\left(A_{1}:A_{k+1}\right),\\
\vdots\\
D(A_{1}:A_{m_{n}+1}:\cdots:A_{N})\geq\sum_{k=m_{n}}^{N-1}D\left(A_{1}:A_{k+1}\right),
\end{array}
\end{eqnarray}
Eq.~(\ref{11}) can be easily verified by summing the above inequalities together.
%So the sum of them is greater than or equal to $\sum_{k=1}^{N-1}D\left(A_{1}:A_{k+1}\right)$,
%then we have
%\begin{eqnarray}
%\triangle D_{G}\leq\triangle D_{C}.
%\label{11}
%\end{eqnarray}

By using the general monogamy relations, we find an interesting trend that if we
divide the $N$-partite system more thoroughly, that is to say, each item of the
$\triangle D_{G}$ contains less parties in the average sense, the $\triangle D_{G}$
will be much larger, the maximum value is $\triangle D_{S}$. On the contrary,
when we divided the $N$-partite system less thoroughly, the $\triangle D_{G}$ will
be much smaller, the minimum value tends to zero. For a better understanding of
the above results, let us consider a quantum system which contains five parties. When
$N=5$, Eq.~(\ref{thm1}) can be reduced to
\begin{eqnarray}
D(A_{1}:\cdots:A_{5})\geq D(A_{1}:A_{2}:A_{3})+D(A_{1}:A_{4}:A_{5}).
\end{eqnarray}
Using the standard monogamy relation, we have
\begin{eqnarray}
D(A_{1}:A_{2}:A_{3})&\geq&D(A_{1}:A_{2})+D(A_{1}:A_{3}),
\nonumber\\
D(A_{1}:A_{4}:A_{5})&\geq&D(A_{1}:A_{4})+D(A_{1}:A_{5}).
\end{eqnarray}
According to the definition (\ref{def1}) and (\ref{def2}),
for this $5$-partite quantum system, the two kinds of monogamy deficits
are as follows:
\begin{eqnarray}
\triangle D_{G}&=&D(A_{1}:\cdots:A_{5})-D(A_{1}:A_{2}:A_{3})-D(A_{1}:A_{4}:A_{5})
\nonumber\\
\triangle D_{S}&=&D(A_{1}:\cdots:A_{5})-\sum_{k=1}^{4}D\left(A_{1}:A_{k+1}\right).
\end{eqnarray}
Obviously, we have $\triangle D_{G}\leq\triangle D_{S}$,
which can be seen as a special case of Eq.~(\ref{11}).

This example clearly shows the intrinsic connection between different kinds
of monogamy deficits. It means that there are different levels of
monogamy deficits.

\section{Another decomposition of the loss of correlation\label{IV}}
From property $(a)$, we know that the loss of correlation for an arbitrary
state $\rho_{A_{1}\cdots A_{N}}$ can be decomposed as the sum of
a series of bipartite ones \cite{key-25}. Since we have this important
identity, an interesting question is whether there is another formula
which connects $D_{\Phi}(A_{1}:\cdots:A_{N})$ and that of subsystems.
In this section, we will consider this question. We will introduce a general form
of this identity
%, then prove it
and present its physical significance.

\begin{thm}
For an arbitrary N-partite system A$_{1}$, A$_{2}$, ... , A$_{N}$,
$D_{\Phi}(A_{1}:\cdots:A_{N})$ satisfies the following identity:
\begin{eqnarray}
&&D_{\Phi}(A_{1}:\cdots:A_{N})\nonumber\\
&=&D_{\Phi}(A_{1}:\cdots:A_{K_{1}})+D_{\Phi}(A_{K_{1}+1}:\cdots:A_{K_{2}})
\nonumber\\
&+&D_{\Phi}(A_{K_{2}+1}:\cdots:A_{K_{3}})+\cdots+D_{\Phi}(A_{K_{N}+1}:\cdots:A_{N})
\nonumber\\
&+&D_{\Phi}(A_{1}\cdots A_{K_{1}}:A_{K_{1}+1}\cdots A_{K_{2}}:%A_{K_{2}+1}\cdots A_{K_{3}}:
\cdots:A_{K_{N}+1}\cdots A_{N}).\ \ \ \
\end{eqnarray}
\end{thm}
This identity is very meaningful because it gives us a lower bound of GQD
\begin{eqnarray}
D(A_{1}:\cdots:A_{N})
\geq D(A_{1}:\cdots:A_{K_{1}})+D(A_{K_{1}+1}:\cdots:A_{K_{2}})
\nonumber\\
+D(A_{K_{2}+1}:\cdots:A_{K_{3}})+\cdots+D(A_{K_{N}+1}:\cdots:A_{N})
\nonumber\\
+D(A_{1}\cdots A_{K_{1}}:A_{K_{1}+1}\cdots A_{K_{2}}:%A_{K_{2}+1}\cdots A_{K_{3}}:
\cdots:A_{K_{N}+1}\cdots A_{N}).\ \
\label{15}
\end{eqnarray}
%equivalent to the sum of the GQDs of its subsystems plus GQD between all these subsystems.
From this inequality, it is easy to show that GQD of an $N$-partite system is always greater
than or equal to the sum of GQDs for its subsystems under any decomposition
\begin{eqnarray}
D(A_{1}:\cdots:A_{N})\geq D(A_{1}:\cdots:A_{K_{1}})
\notag\\
+\cdots+D(A_{K_{N}+1}:\cdots:A_{N}).
\end{eqnarray}
For the same reason, the GQD is also not less than GQD between all its subsystems under any decomposition
\begin{eqnarray}
D(A_{1}:\cdots:A_{N})\geq D(A_{1}\cdots A_{K_{1}}:\cdots:A_{K_{N}+1}\cdots A_{N}).
\end{eqnarray}

Furthermore, using the above result, we give a proposition:%more general conclusions which
%has the following form:
\begin{prop}
For an arbitrary N-partite system A$_{1}$, A$_{2}$, ... , A$_{N}$ and an arbitrary nonnegative
integer $n$, the inequality:
\begin{widetext}
\begin{eqnarray}
[D(A_{1}:\cdots:A_{N})]^{n}
&\geq&[D(A_{1}:\cdots:A_{K_{1}})]^{n}+[D(A_{K_{1}+1}:\cdots:A_{K_{2}})]^{n}
+[D(A_{K_{2}+1}:\cdots:A_{K_{3}})]^{n}+\cdots+[D(A_{K_{N}+1}:\cdots:A_{N})]^{n}
\nonumber\\
&+&[D(A_{1}\cdots A_{K_{1}}:A_{K_{1}+1}\cdots A_{K_{2}}:A_{K_{2}+1}\cdots A_{K_{3}}:\cdots:A_{K_{N}+1}\cdots A_{N})]^{n}
\end{eqnarray}
\end{widetext}
holds .
\end{prop}
%This formula is very interesting, since it tells us that GQD of an $N$-partite system
%to the power of $n$ is always not less than the sum of GQDs for its subsystems to the
%power of $n$ the plus GQD between all these subsystems to the power of $n$.
It's worth noting that GQD to the power of $n$ can be seen as a multipartite quantum
correlation. In this sense, the multipartite quantum correlation for
an $N$-partite system is always not less than the sum of multipartite quantum
correlations for its subsystems plus the quantum correlation between all these subsystems.
Let us consider a simple example to understand the above results better. When $N=4$,
Eq.~(\ref{15}) reduces to
\begin{eqnarray}
D(A_{1}:A_{2}:A_{3}:A_{4})\geq D(A_{1}:A_{2})+D(A_{3}:A_{4})\nonumber\\
+D(A_{1}A_{2}:A_{3}A_{4}).\ \
\label{w}
\end{eqnarray}
%That is, GQD of this $4$-partite system can be decomposed into the sum of its $2$-partite
%subsystems plus the GQD between the two subsystems.
%
Since GQD is always non-negative for arbitrary states, we have
\begin{eqnarray}
D(A_{1}:A_{2}:A_{3}:A_{4})\geq D(A_{1}:A_{2})+D(A_{3}:A_{4})
\notag\\
D(A_{1}:A_{2}:A_{3}:A_{4})\geq D(A_{1}A_{2}:A_{3}A_{4}).
\end{eqnarray}
According to the above general conclusion, the following relation always holds
\begin{eqnarray}
[D(A_{1}:A_{2}:A_{3}:A_{4})]^{n}\geq[D(A_{1}:A_{2})]^{n}+[D(A_{3}:A_{4})]^{n}
\notag\\
+[D(A_{1}A_{2}:A_{3}A_{4})]^{n}.\ \ \ \
\end{eqnarray}
This formula can be seen as an extension of the equation (19), when $n=1$, it reduces to Eq.~(\ref{w}). If we consider the GQD to the power of n as a multipartite quantum correlation, this result tells us that the multipartite quantum correlation is always greater than or equal to the sum of multipartite quantum correlations for its subsystems plus the quantum correlation between all these subsystems.

\section{The second class of monogamy inequality of GQD \label{V}}
As far as we know, GQD obeys the standard monogamy relation under the
condition that the bipartite GQDs do not increase when
some subsystems are discarded \cite{key-25}.
%This monogamy relation reflects the distribution
%property of GQD from one aspect.
In this section, to further understand
the distribution of GQD, we will introduce another trade-off
inequality which is also upper bounded by the GQD of a multipartite system.
It reflects how GQD is distributed in the multipartite quantum system
from another aspect, which can also be seen as a generalized monogamy
inequality. We call it the second class of monogamy inequality of
GQD. %We first introduce a general form of this inequality, than we
%prove it and give its physical interpretation.

\begin{thm}
For an arbitrary $N$-partite system $A_{1}$, $A_{2}$, ... , $A_{N}$ and integer $K$ with $1<K<N$,
GQD of it satisfies the second class of monogamy inequality which
has the following form:
\begin{eqnarray}
D(A_{1}:\cdots:A_{N})&-&\sum_{i=1}^{N-K}D(A_{i}\cdots A_{i+K-1}:A_{i+K})
\notag\\
&\geq& D(A_{1}:A_{2}:\cdots:A_{K}),
\label{22}
\end{eqnarray}
provided that the bipartite GQDs do not increase under the discard of subsystems.
\end{thm}

Since the general form is complex, in order to make its physical significance
more clearly, we consider some special cases.
First of all, let us show the case that $N=4$, $K=2$, then Eq.~(\ref{22}) reduces to
\begin{eqnarray}
D(A_{1}:A_{2})\leq D(A_{1}:\cdots:A_{4})-D(A_{1}A_{2}:A_{3})\nonumber\\
-D(A_{2}A_{3}:A_{4}).
\end{eqnarray}

In Ref.~\cite{key-25}, the first monogamy relation Eq.~(\ref{5}) holds provided
that the bipartite GQDs $D\left(A_{1}\cdots A_{k}:A_{k+1}\right)$ do not increase
under discard of subsystems, that is,
$D\left(A_{1}\cdots A_{k}:A_{k+1}\right)\geq D\left(A_{1}:A_{k+1}\right)$.
Analogously, when $K=1$, Eq.~(\ref{22}) has an interesting special case %we can perform the second monogamy relation under the similar condition.
%Furthermore,,,
and it degenerates into the following form:
\begin{eqnarray}
D(A_{1}:\cdots:A_{N})\geq D(A_{1}:A_{2})+D(A_{2}:A_{3})\nonumber\\
+\cdots+D(A_{N-1}:A_{N}),
\end{eqnarray}
which is performed as  the second monogamy relation under the similar condition.
The above inequality shows that GQD of an $N$-partite system
is greater than or equal to the sum of GQDs between two nearest
neighbor particles.

Unlike this, the standard monogamy relation tells
us that the GQD of an $N$-partite system is always greater than or equal
to the sum of GQDs between one particle and each of remaining $(N-1)$-particles.
The physical meaning of these two monogamy relations are quite different.
For example, when we consider the $4$-partite system, the standard monogamy
relation is as follows:
\begin{equation}
D(A_{1}:\cdots:A_{4})\geq D(A_{1}:A_{2})+D(A_{1}:A_{3})+D(A_{1}:A_{4}).\label{tree}
\end{equation}
On the other hand, the second monogamy relation has the following
form:
\begin{equation}
D(A_{1}:\cdots:A_{4})\geq D(A_{1}:A_{2})+D(A_{2}:A_{3})+D(A_{3}:A_{4}).\label{chain}
\end{equation}
\begin{figure}[t]
 \centering
\includegraphics[width=0.45\textwidth]{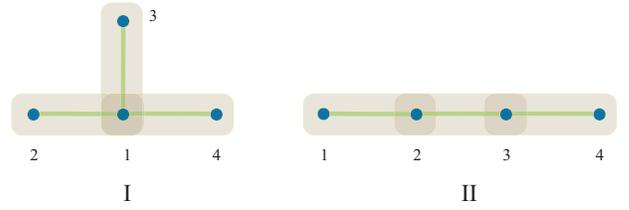}\\
\caption{(color online). Two different configurations of monogamy relations.
``Tree structure'' (I) is for Eq.~(\ref{tree}); and (II) as a ``linked list'' is for Eq.~(\ref{chain}).
\label{fig:1}}
\end{figure}
\begin{figure}[t]
 \centering
\includegraphics[width=0.42\textwidth]{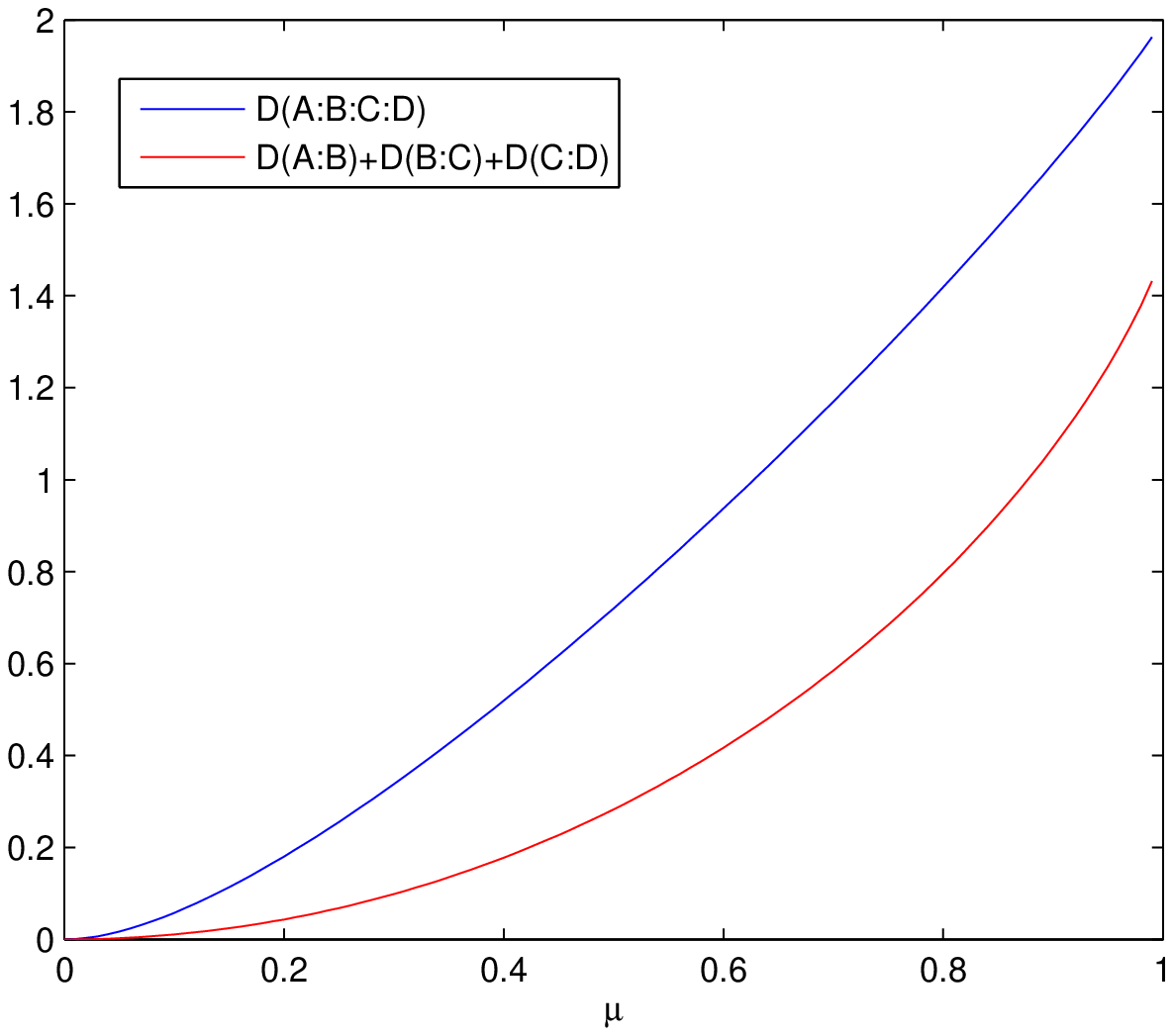}\\
\caption{(color online). $D\left(A:B:C:D\right)$ and
$D\left(A:B\right)+D\left(B:C\right)+D\left(C:D\right)$ v.s. $\mu$.}
\label{fig:2}
\end{figure}

It is easy to show that these two monogamy relations are very different,
not only in the physical meaning, but also in the structures of graph theory.
%In fact, they have different topologies.
A schematic description comparing these two monogamy relations is displayed
in Fig.~\ref{fig:1}.
%From this figure,
%we know that the topology of standard monogamy relation is $T$-shaped.
%On the other hand, the second monogamy relation has a linear topology.
%In other words,
From this figure, %using the language of graph theory,
the standard monogamy relation relates to a ``tree structure'', the second monogamy
relation relates to a ``linked list''. Therefore, these two monogamy relations are not equivalent.
%another by using any method.
It's worth noting that when we consider the quantum system which contains
more parties, there will be more different structures for these
two monogamy relations. The number of different structures increases
very rapidly with growing $N$. This fact tells us that the monogamy
structures of GQD are more than people have considered in the
previous literature.

For instance, we consider a family of states as follows,
$\rho={(1-\mu)}I^{\otimes N}/{2^{N}}+\mu|W(N)\rangle\langle W(N)|$,
where $I$ is the identity and $\mu\in\left[0,1\right]$.
Note that $|W(N)\rangle$ is the $N$-partite W state
$(|10\cdots0\rangle+|01\cdots0\rangle+\cdots+|00\cdots1\rangle)/{\sqrt{N}}$.
In Fig.~\ref{fig:2}, $D\left(A:B:C:D\right)$ and
$D\left(A:B\right)+D\left(B:C\right)+D\left(C:D\right)$ are plotted as
a function of $\mu$. This figure shows that both two quantities increase
as $\mu$ grows, $D\left(A:B:C:D\right)$ is always greater than
or equal to the $D\left(A:B\right)+D\left(B:C\right)+D\left(C:D\right)$.
%, so
%the second monogamy inequality always hold for this case.
Moreover, the difference between these two quantities first increases and
then decreases as $\mu$ increases. That is to say, the non-nearest neighbor correlation first increases then decreases as the state becomes more closer to the $N$-partite W state with increasing $\mu$. For appropriate $\mu$, this non-nearest neighbor correlation reaches a maximum. It tells us that this correlation reaches a maximum for a part of mixed state, neither the completely mixed state nor the pure state. In fact, it is corresponding to the "residual GQD" that we will considered in next section.

\begin{figure}[t]
 \centering
\includegraphics[width=0.42\textwidth]{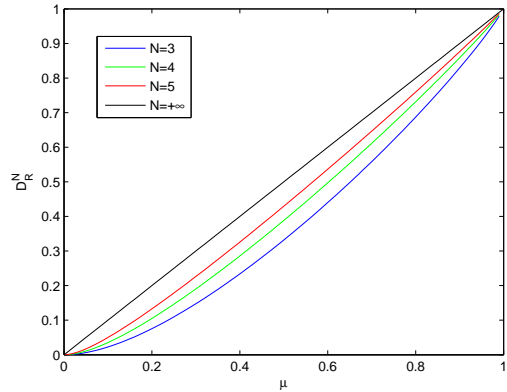}\\
\caption{(color online). $D_{R}^{N}$ of Werner-GHZ state against $\mu$ for different $N$.}
\label{fig:3}
\end{figure}
\begin{figure}[t]
 \centering
\includegraphics[width=0.42\textwidth]{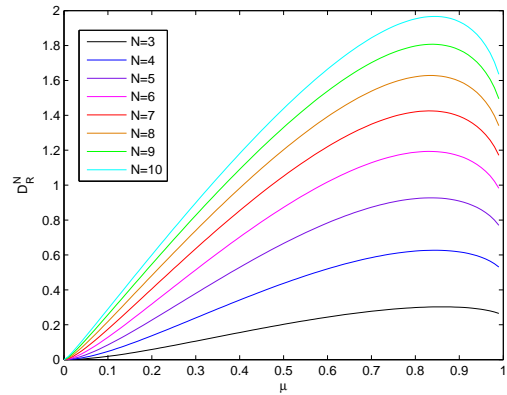}\\
\caption{(color online). $D_{R}^{N}$ of mixed W state against $\mu$ for different $N$.}
\label{fig:4}
\end{figure}

\section{Residual GQD\label{VI}}

Then, similar to the definition of tangle as a measure of residual multipartite
entanglement, we can define the residual GQD corresponding to the second
monogamy relation,
\begin{eqnarray}
D_{R}^{N}\equiv D(A_{1}:\cdots:A_{N})-\sum_{K=1}^{N-1}D(A_{K}:A_{K+1}).\label{DR}
\end{eqnarray}
It is a measure for residual multipartite quantum correlations, namely, contributions
to quantum correlations beyond pairwise GQD. This measure of residual
multipartite quantum correlations describes the total quantum correlation except for
all the nearest neighbor interaction of quantum correlations.
%Using the same method as last section, we can get
%\begin{eqnarray}
%D_{R}^{\left(N\right)}-D_{R}^{\left(N-1\right)}=I((A_{1}\cdots A_{N-2})A_{N}\mid A_{N-1})
%\notag\\
%-I_{\Phi}((A_{1}\cdots A_{N-2})A_{N}\mid A_{N-1}).
%\end{eqnarray}

%For a better understanding of the monogamy deficit of GQD, we consider the following states.
%First, we consider $N$-qubit ($N\geq2$) Werner-GHZ state \cite{key-29} in Fig.~\ref{fig:3},
%$\rho=\frac{(1-\mu)I^{\bigotimes N}}{2^{N}}+\mu|\psi\rangle\langle\psi|$,
%where, I is $2\times2$ identity operator, $\mu\in\left[0,1\right]$, $|\psi\rangle$ is the $N$-qubit GHZ state
%$|\psi\rangle=\frac{1}{\sqrt{2}}\left(|00\ldots0\rangle+|11\ldots1\rangle\right)$.
%From this figure, we can see that the residual GQD is always greater than or equal to zero for the
%Werner-GHZ state. For more detail, $D_{R}^{\left(N\right)}$  increases with increasing $N$, when
%$N\rightarrow\infty$, the $D_{R}^{\left(N\right)}$ approaches to the maximum value $\mu$.
%Furthermore, it is obvious that the $D_{R}^{\left(N\right)}-D_{R}^{\left(N-1\right)}$ decreases
%with increasing $N$, when $N\rightarrow\infty$, $D_{R}^{\left(N\right)}-D_{R}^{\left(N-1\right)}$
%approaches to zero. That is to say, for this state, the interrogated conditional mutual information
%with respect to $A_{1}$ is always less than or equal to the unmeasured conditional mutual information
%with respect to $A_{1}$, the monogamy inequality is always hold.

In order to better understand the residual GQD, let us consider two examples.
First, we consider a $N$-qubit ($N\geq2$) Werner-GHZ state \cite{key-29},
$\rho={(1-\mu)I^{\otimes N}}/{2^{N}}+\mu|\psi\rangle\langle\psi|$,
where  $|\psi\rangle$ is the $N$-qubit GHZ state
$|\psi\rangle=\left(|00\ldots0\rangle+|11\ldots1\rangle\right)/{\sqrt{2}}$. In Fig.~\ref{fig:3},
the residual GQD $D_{R}^{N}$ of Werner-GHZ state is plotted as a function of $\mu$.
From this figure, we can see that the residual GQD is always greater than or equal to zero.
For more details, $D_{R}^{N}$  increases with an increasing $N$; when
$N\rightarrow\infty$, $D_{R}^{N}$ approaches to the maximum value $\mu$.
Furthermore, it is obvious that $D_{R}^{N}-D_{R}^{N-1}$ decreases
with an increasing $N$; when $N\rightarrow\infty$, $D_{R}^{N}-D_{R}^{N-1}$
approaches to zero.

Next, we investigate the state considered in the former section,
$\rho={(1-\mu)}I^{\otimes N}/{2^{N}}+\mu|W(N)\rangle\langle W(N)|$.
%,
%note that $|W(N)\rangle$ is the $N$-partite W state
%$\frac{1}{\sqrt{N}}(|10\cdots0\rangle+|01\cdots0\rangle+\cdots+|00\cdots1\rangle)$.
In Fig.~\ref{fig:4}, the residual GQD:
\begin{widetext}
\begin{eqnarray}
D_{R}^{N}
&=&\left[\left(N-1\right)\frac{1-\mu}{2^{N}}\log_{2}\left(\frac{1-\mu}{2^{N}}\right)
-N\left(\frac{1-\mu}{2^{N}}+\frac{\mu}{N}\right)\log_{2}\left(\frac{1-\mu}{2^{N}}+\frac{\mu}{N}\right)
+\left(\frac{1-\mu}{2^{N}}+{\mu}\right)\log_{2}\left(\frac{1-\mu}{2^{N}}+{\mu}\right)\right]\nonumber\\
%&-&(N-1)\left[\frac{1-\mu}{4}\log_{2}\left(\frac{1-\mu}{4}\right)
%+\left(\frac{1-\mu}{4}+\frac{N-2}{N}\mu\right)\log_{2}\left(\frac{1-\mu}{4}+\frac{N-2}{N}\mu\right)
%+2\left(\frac{1-\mu}{4}+\frac{\mu}{N}\right)\log_{2}\left(\frac{1-\mu}{4}+\frac{\mu}{N}\right)\right]\nonumber\\
&+&(N-1)\left[\left(\frac{1-\mu}{4}\right)\log_{2}\left(\frac{1-\mu}{4}\right)
+\left(\frac{1-\mu}{4}+\frac{2\mu}{N}\right)\log_{2}\left(\frac{1-\mu}{4}+\frac{2\mu}{N}\right)
-2\left(\frac{1-\mu}{4}+\frac{\mu}{N}\right)\log_{2}\left(\frac{1-\mu}{4}+\frac{\mu}{N}\right)\right]
%+\left(\frac{1-\mu}{4}+\frac{N-2}{N}\mu\right)\log_{2}\left(\frac{1-\mu}{4}+\frac{N-2}{N}\mu\right)\right]
\end{eqnarray}
\end{widetext}
is plotted as a function of $\mu$. This
figure shows that the $D_{R}^{N}$ is always greater than or equal to zero for
different $N$, so the second monogamy inequality always holds for this state. It's worth noting
that the $D_{R}^{N}$ always first increases and then decreases for different $N$,
which is different from the behavior of the $N$-qubit Werner-GHZ state. Similar as the case in last
section, the $D_{R}^{N}$ is always greater than or equal to $D_{R}^{N-1}$.
For more details, it is obvious that when $N$ increases, the
$D_{R}^{N}-D_{R}^{N-1}$ decreases.
%In fact, when $N\rightarrow\infty$,
%the $D_{R}^{\left(N\right)}$ becomes a straight line,
When $\mu\rightarrow1$,
$D_{R}^{N}$ %approaches to $\mbox{\ensuremath{\log_{2}}}N$
%that
is divergent for large $N$.
%That is to say, for this case, the interrogated conditional mutual
%information with respect to $A_{K}$ is always less than or equal to the unmeasured conditional
%mutual information with respect to $A_{K}$, the second monogamy inequality always holds.

\section{Global quantum discord and quantum phase transitions }\label{VII}

It is known that quantum correlations such as entanglement \cite{amico,key-30,key-31,key-32,fan1,fan2}, differential
local convertibility \cite{p3}, non-locality \cite{key-33} and bipartite quantum discord \cite{key-34}
can be applied to study the quantum phase transitions. The behavior
of bipartite and global correlations in the quantum spin chains, especially for Ising model, has attracted
considerable interest. Recently, it is worth noting that the behavior of global quantum discord of
a finite-size transverse field Ising model near its critical point has been studied \cite{key-35}.
Next, we also use Ising model with transverse filed as an example to show the
application of the GQD. Let us consider a one dimensional Hamiltonian with
periodic boundary conditions as follows:
\begin{equation}
\hat{H_{I}}=-J\sum_{i=1}^{L}\hat{\sigma}_{i}^{z}\hat{\sigma}_{i+1}^{z}+B\sum_{i=1}^{L}\hat{\sigma}_{i}^{x},
\end{equation}
where we set the condition $L+1\equiv1$. According to previous literature \cite{key-31}, in the limit
$B/J\rightarrow0$, the ground state of this model is locally equivalent to an $L$-spin GHZ state.  As $B$
increases, in the thermodynamic limit, the system undergoes a quantum phase transition at $B/J=1$.

In this section, in line with \cite{key-35}, we will study (i) the total GQD for this $N$-partite spin system,
(ii) sum of all the nearest neighbor bipartite GQDs, and (iii) the residual GQD as $B/J$ changes at $T=0$ and $T=0.1$. We will
show that the sum of all the nearest neighbor bipartite GQDs is more effective and accurate for signaling the
critical point of quantum phase transitions and the non-local correlations, such as residual GQD also plays an
interesting role in this model.

\begin{figure}[t]
\centering
\subfigure[]{\includegraphics[width=0.33\textwidth]{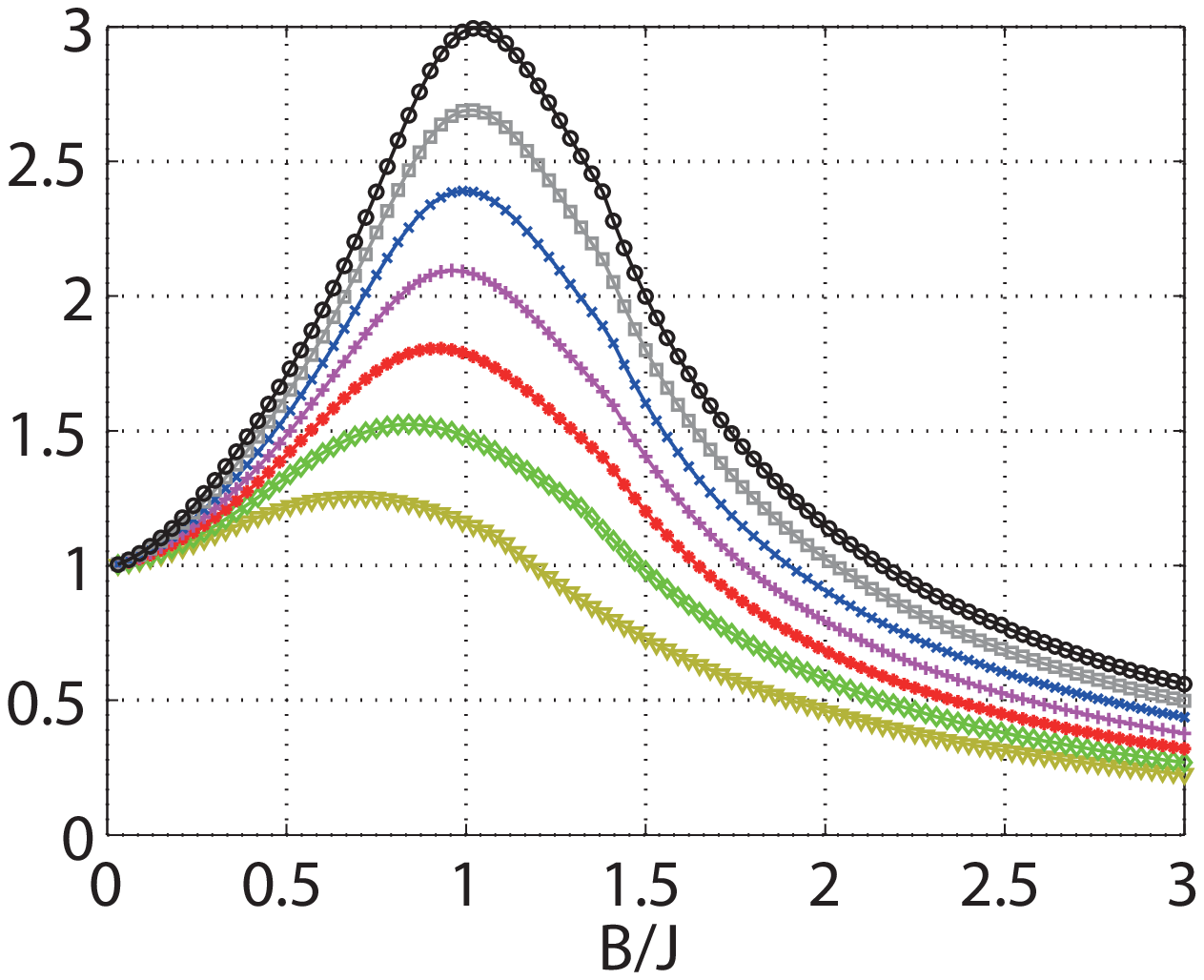}\label{f51}}\\
\subfigure[]{\includegraphics[width=0.33\textwidth]{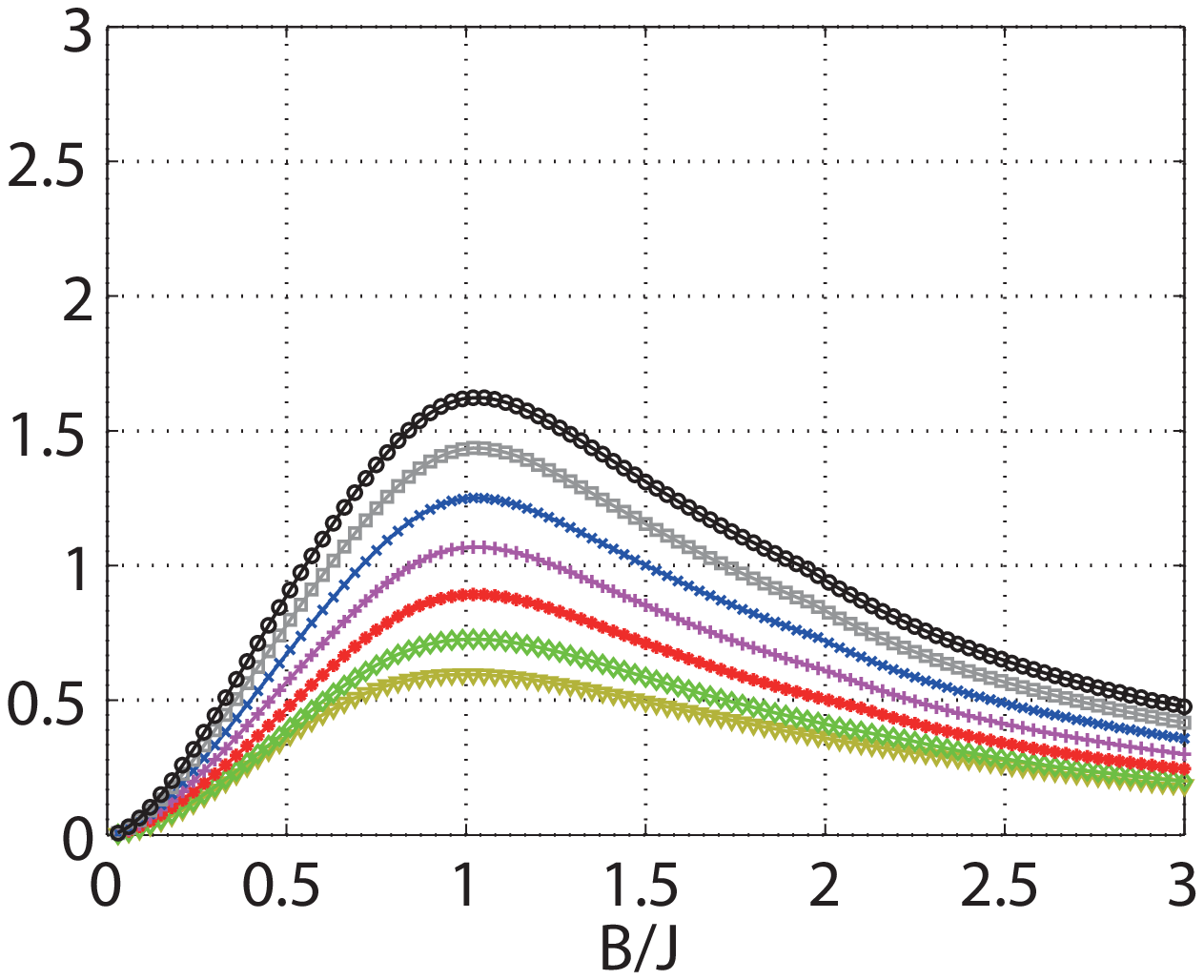}\label{f52}}\\
\subfigure[]{\includegraphics[width=0.33\textwidth]{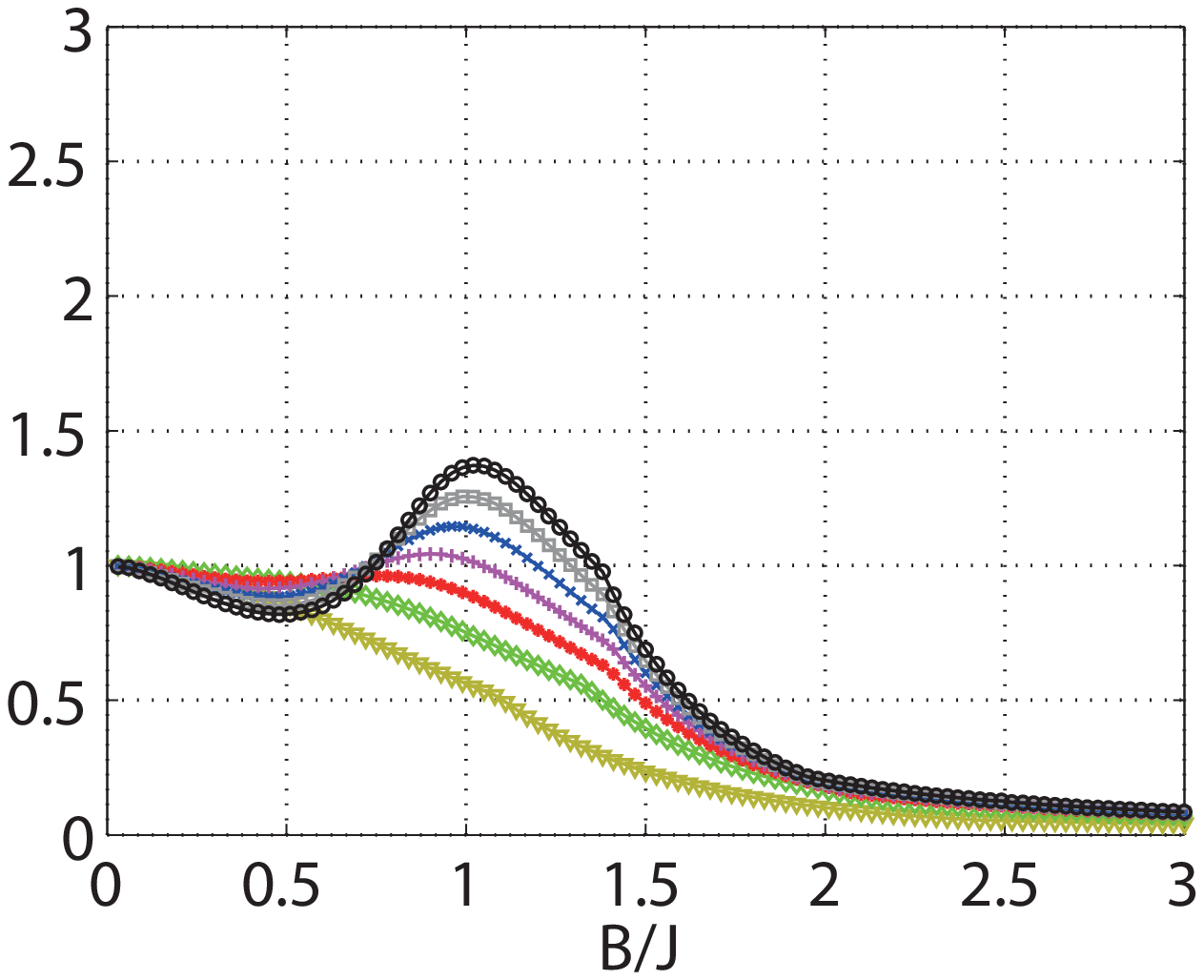}\label{f53}}\\
\caption{(color online). (a) GQD for the Ising model at zero temperature. From bottom to top curve,
$L$ goes from $3$ to $9$ spins. At $B\rightarrow0$ GQD =1 since the ground state
is a GHZ state. In the paramagnetic configuration ( $B/J\gg1$ ),
GQD goes to zero. (b) The sum of all the nearest neighbor bipartite GQDs for the Ising model
at zero temperature. From bottom to top curve, $L$ goes from $3$ to
$9$ spins. At $B\rightarrow0$, we have $\sum_{i=1}^{L-1}D\left(A_{i}:A_{i+1}\right)=0$.
In the paramagnetic configuration ( $B/J\gg1$ ), $\sum_{i=1}^{L-1}$
D $\left(A_{i}:A_{i+1}\right)$ goes to zero together with total GQD. (c) The residual GQD for the Ising model at zero temperature. From bottom
to top curve, $L$ goes from $3$ to $9$ spins. At $B\rightarrow0$ residual GQD
equal to 1. In the region that $B/J\gg1$, the residual GQD goes to
zero faster than total GQD.}\label{fig:5}
\end{figure}

First, we consider the case of zero temperature. In order to calculate
the global quantum correlations mentioned above, we first reformulate
GQD as \cite{key-35}:
\begin{eqnarray}
D(A_{1}:\cdots:A_{L})&=&\underset{\{ \hat{\Pi^{k}}\} }{\min}\left\{ \sum_{j=1}^{L}\sum_{l=0}^{1}{\tilde{\rho}_{j}^{ll}}\log_{2}\tilde{\rho}_{j}^{ll}-\sum_{k=0}^{2^{L}-1}\tilde{\rho}_{T}^{kk}\log_{2}\tilde{\rho}_{T}^{kk}\right\} \nonumber\\
&+&\sum_{j=1}^{L}S(\rho_{j})-S(\rho_{T})
\end{eqnarray}
with $\tilde{\rho}_{T}^{kk}=\langle\bm{k}|\hat{{R}}^{\dag}\rho_{T}\hat{{R}}|\bm{k}\rangle$ and
$\tilde{\rho}_{j}^{ll}=\langle l|\hat{{R}}_{j}^{\dag}\rho_{j}\hat{{R}}_{j}| l\rangle$, where
$\hat{\Pi}^{k}=\hat{R}|\bm{ k}\rangle\langle \bm{k}|\hat{R}^{\dag}$ are the multi-qubit
projective operators. Here $\left\{ |\bm{k}\rangle\right\}$ are separable eigenstates of
$\bigotimes_{j=1}^{L}\hat{\sigma}_{j}^{z}$, %with $\sigma_{j}^{q}$  the $q=x,y,z$ Pauli operator
and $\hat{R}$ is a local $L$-qubit rotation: $\hat{R}=\bigotimes_{j=1}^{L}\hat{R}_{j}(\theta_{j},\phi_{j})$
with $\hat{R_{j}}(\theta_{j},\phi_{j})=\cos\theta_{j}\hat{I}+i\sin\theta_{j}\cos\phi_{j}\hat{\sigma}_{y}+i\sin\theta_{j}\sin\phi_{j}\hat{\sigma}_{x}$
%the rotation operator (of angles $\theta$ and $\phi_{j}$)
acting on the $j$-th qubit.
%This formula greatly reduces the computational efforts needed to evaluate
%GQD.
Considering the symmetries of this model, the GQD is completely independent
on the set of $\phi_{j}$ angles, so we only need to optimize the
$\theta_{j}$. Similar as the literature \cite{key-35}, the optimal $\theta_{j}$
all take the same value $\bar{\theta}$, which depends on the magnetic
field.
%We have $\bar{\theta}=0\ \left[\pi/4\right]$ for small {[}large{]} values of $B$.

Fig.~\ref{f51} shows the GQD as a function of the ratio $B/J$ between the
magnetic field intensity and the Ising interaction constant when $T=0$.
We study rings with $L=3,\ldots,9$. GQD curves share the same value $1$ as
$B\rightarrow0$, since the ground state of our model is an $L$-spin GHZ state.
When $B/J$ tends to $1$, the GQD increases reaching a maximum at different
positions depending on size $L$ of the system. In the paramagnetic phase achieved for
$B/J\gg1$, all the spins align along the direction of the magnetic field, so that global
discord disappears.
We find that the height and position of  the maximum of GQD is varies in accordance
with the system size $L$.

%However, the GQD undergoes significant changes which result in the
%appearance of a maximum, whose height and position is a clear function of the
%system size $L$.

Fig.~\ref{f52} shows the sum of all the nearest neighbor bipartite GQDs:
$D\left(A_{1}:A_{2}\right)+D\left(A_{2}:A_{3}\right)+\cdots+D\left(A_{L-1}:A_{L}\right)$
as a function of the ratio $B/J$ when $T=0$. We still study rings with
$L=3,\ldots,9$, whose curves share the same value $0$ at $B\rightarrow0$. As
$B/J$ tends to $1$, $\sum_{i=1}^{L-1} D \left(A_{i}:A_{i+1}\right)$
reaches a maximum at nearly $B/J=1$, which almost independent on
the size of the system. It shows that the behavior of the sum of all
the nearest neighbor bipartite GQDs is quite different with total
GQD, and is more suited to be used to describe the quantum phase
transitions in this model. It is because that our Hamiltonian only contains the nearest-neighbor
interactions. In the region that $B/J\gg1$, similar as the total
GQD, the sum of all the nearest neighbor bipartite GQDs also disappears.
%It is obvious that when $B/J$ tends to $1$, the difference between
%neighboring lines reaching a maximum, which also closely related to
%the quantum phase transitions.

Fig.~\ref{f53} shows the residual GQD: $D_{R}^{L}$ (see Eq.~(\ref{DR})) as a function
of the ratio $B/J$ when $T=0$. We still study rings with $L=3,\ldots,9$,
whose residual GQD curves share the same value $1$ at $B\rightarrow0$ since the
sum of all the nearest neighbor bipartite GQDs disappears at this point.
The behavior of residual GQD is very sensitive to the system's size $L$. For $L=3,4$, the residual GQD is monotonically decreasing.
When the system's size $L$ increases, the residual GQD is not monotonous.
As $B/J$ is around $1$, it  can also be used to characterize the
quantum phase transitions in this model when we consider the appropriate
system's size $L$. This fact tells us that for the large system, the
non-nearest-neighbor correlations also play an important role in the
quantum phase transitions although our Hamiltonian only contains the
nearest neighbor interactions. This phenomenon reflects that the nearest
neighbor interactions can also create long-range correlations.
In
the region that $B/J\gg1$, the residual GQD tends to $0$ faster than
%the total GQD and
the sum of all the nearest neighbor bipartite GQDs.
It shows that when we consider the large magnetic field, the long-range
correlations disappear firstly.

%Combining the above three figures, we find an interesting relation
%among them.
Comparing these three figures in Fig.~\ref{fig:5}, we find that the sum of all the nearest neighbor bipartite GQDs
gives the best criteria for the critical point of the quantum phase transitions.
Since the total GQD is the sum of residual GQD together with the sum of all the nearest neighbor bipartite GQDs,
the shift of the maximum point of total GQD away from the critical point may be due to the shift of the maximum
of residual GQD.

\begin{figure}[t]
\centering
\subfigure[]{\includegraphics[width=0.33\textwidth]{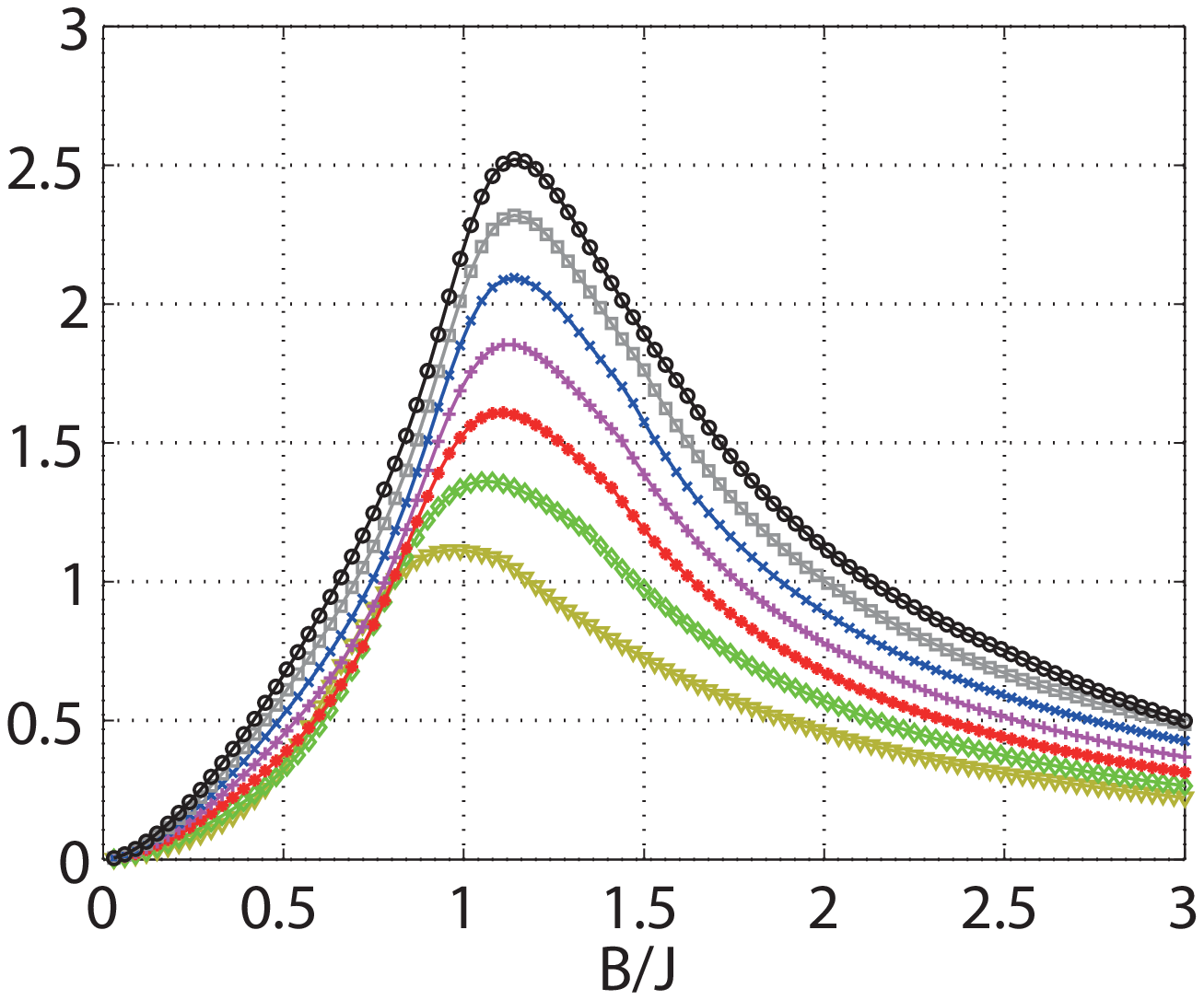}\label{f61}}\\
\subfigure[]{\includegraphics[width=0.33\textwidth]{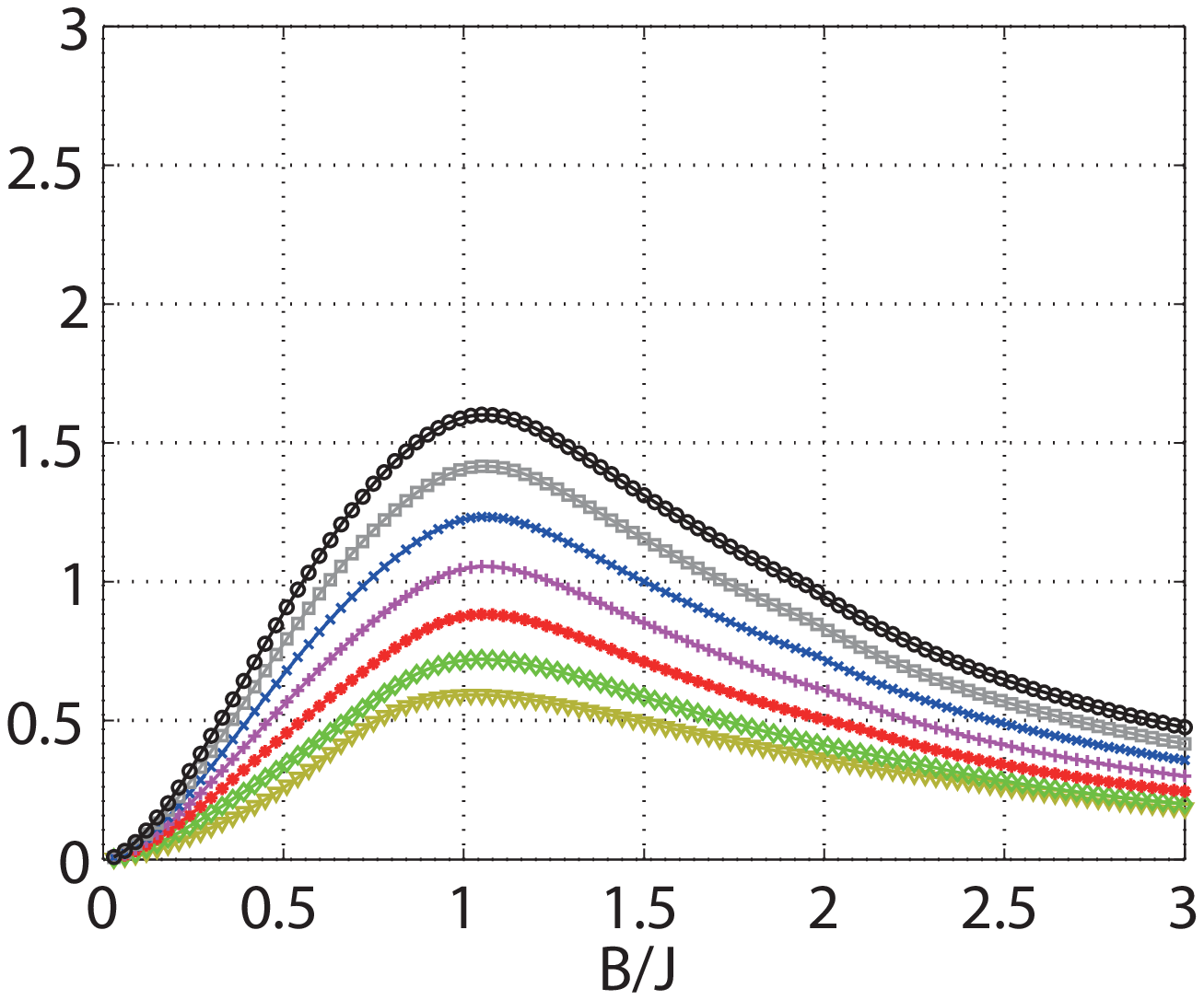}\label{f62}}\\
\caption{(color online). (a) GQD for the Ising rings containing from $3$ to $9$ spins at $T=0.1$.
GQD is null at zero magnetic field. For increasing temperatures the
maximum values of the curves decrease. (b) The sum of all the nearest neighbor bipartite GQDs for the Ising rings
containing from $3$ to $9$ spins at $T=0.1$. $\sum_{i=1}^{L-1}$ D $\left(A_{i}:A_{i+1}\right)$
is null at $B=0$. In the region that $B/J\gg1$, $\sum_{i=1}^{L-1}$
D $\left(A_{i}:A_{i+1}\right)$ goes to zero together with total GQD.}\label{fig:6}
\end{figure}

Next, we consider the case that $T\neq0$. Similar as \cite{key-35}, we take the Gibbs state as our thermal state:
\begin{equation}
\rho_{T}=\frac{e^{-\hat{H}/T}}{\mathcal{Z}}
\end{equation}
with $\hat{H}_{I}$ the Hamiltonian describing the interaction, $T$ the
effective temperature, and $\mathcal{Z}=\textrm{Tr}[\exp({-\hat{H}_{I}/T})]$ the partition
function.

We consider the behavior of total GQD and the sum of all the nearest
neighbor bipartite GQDs with effective temperature  $T=0.1$ in Fig.~\ref{f61} and Fig.~\ref{f62},
respectively. Fig.~\ref{f61} shows the GQD as a function of the ratio
$B/J$ when $T=0.1$. It's worth noting that at non-zero temperature,
the quantum correlations presented in the ground state at $B=0$
are destroyed since as $B\rightarrow0$ the ground and first excited
states approach degeneracy. It is obvious that the height of the curves
decreases with increasing temperature and there is a right shift in
the maxima of each curve. %The optimal angles for the GQD of thermal
%states are the same as those for zero temperature.

Fig.~\ref{f62} shows the sum of all the nearest neighbor bipartite GQDs as
a function of the ratio $B/J$ when $T=0.1$. Comparing Fig.~\ref{f62} with Fig.~\ref{f52},
it is amazing that the behavior of $\sum_{i=1}^{L-1} D \left(A_{i}:A_{i+1}\right)$
is almost independent of the effective temperature $T$. That is to
say, the sum of all the nearest neighbor bipartite GQDs is more suited
to be used to describe the quantum phase transition in our model both
for zero temperature and non-zero effective temperature. We remark that
the residual GQD can be negative for non-zero temperature, so it is
no longer a well-defined quantum correlation in this case. This fact
reflects that at non-zero temperature the non-nearest-neighbor correlations
are destroyed.

\section{conclusions and discussion }
In summary, we have introduced a series of generalized monogamy relations
of GQD for an $N$-partite system. Remarkably, these monogamy relations
hold for general states whose bipartite GQDs is non-increasing in
discarding of subsystems. Using the decomposition property of GQD, we
provide a family of monogamy inequalities which can be considered as an
extension of the standard monogamy inequality. We have proved that they
hold under the similar condition as that for standard monogamy relation
and shown that there is an intrinsic connection between them.
%By using the definition
%of GQD and the decomposition property of multipartite mutual information,
We also demonstrate that GQD of an $N$-partite system is not less than the sum of
GQDs of its subsystems plus GQD between all its subsystems under any decomposition.
%find an important identity that tells us that GQD for an
%$N$-partite system is equivalent to the sum of GQDs for its subsystems
%plus GQD between all these subsystems. From this identity, it
%is easy to show that GQD of an $N$-partite system is always not less than
%the sum of GQDs for its subsystems or GQD between all its subsystems under
%any decomposition.
Furthermore, we have provided the second class of monogamy
inequality which is also upper bounded by GQD of a multipartite system. It reflects
how GQD is distributed in the multipartite quantum system from a new aspect.
In particular, when $K=1$, one of the special case of the second class
of monogamy inequality is obtained and shows that the GQD of an $N$-partite system
is greater than or equal to the sum of GQDs between two nearest
neighbor particles. Its physical meaning is quite different from the
standard monogamy relation. Last but not least,
%we provided two kinds of
%monogamy deficit of GQD and give the necessary and sufficient conditions
%for the two monogamy inequality hold.
we provide the residual GQD corresponding to the second monogamy inequality and
study its properties by considering two typical states.

More importantly, we demonstrate an interesting application of the sum of all the nearest neighbor
bipartite GQDs and residual GQD . By considering their behavior
in the transverse field Ising model, we find that both of them can
be used to characterize the quantum phase transitions at zero temperature. It is worth
noting that the sum of all the nearest neighbor bipartite GQDs is
more suited to be used to describe the quantum phase transition in
our model both for zero temperature and non-zero effective temperature
case. This result is superior to the the results obtained in the previous
literature \cite{key-35}. That is to say, we provide a new and more effective
way to characterize the quantum phase transitions in this model.
%All these results can be
%generalized to q-GQD for the appropriate value of q.

We believe that our result provide a useful method in understanding the distribution
property of GQD in multipartite quantum systems.
The introduced quantities not only play a fundamental role in the
quantum information processing, but also can be applied to physical models of many-body quantum systems.

\begin{acknowledgments}
We thank Yan-Kui Bai,  Yu Zeng and Xian-Xin Wu for valuable discussions. This work is supported
by ``973'' program (2010CB922904), NSFC (11075126, 11031005,
11175248) and NWU graduate student innovation funded YZZ12083.
\end{acknowledgments}

\begin{widetext}
\section*{APPENDIX}
\renewcommand{\theequation}{A1-\arabic{equation}}
% redefine the command that creates the equation no.
\setcounter{equation}{0}  % reset counter
\subsection*{A1.~Proof of Theorem 1}
\begin{proof}
In order to simplify our proof, we first consider a special case as
follows:
\begin{eqnarray}
D(A_{1}:\cdots:A_{N})\geq D(A_{1}:A_{2}:\cdots:A_{m})
+D(A_{1}:A_{m+1}:\cdots:A_{N}).
\end{eqnarray}
According to the definition of GQD, we have
$D(A_{1}:\cdots:A_{N})=\underset{\Phi}{\min}D_{\Phi}(A_{1}:\cdots:A_{N})$.
Using property $\left(a\right)$, we can rewrite
$D_{\Phi}(A_{1}:\cdots:A_{N})$ as following form:
\begin{eqnarray}
D_{\Phi}(A_{1}:\cdots:A_{N})&=&\sum_{k=1}^{N-1}D_{\Phi}\left(A_{1}\cdots A_{k}:A_{k+1}\right)
=\sum_{k=1}^{m-1}D_{\Phi}\left(A_{1}\cdots A_{k}:A_{k+1}\right)+\sum_{k=m}^{N-1}D_{\Phi}\left(A_{1}\cdots A_{k}:A_{k+1}\right)
\nonumber\\
&=&D_{\Phi}(A_{1}:\cdots:A_{m})+\sum_{k=m}^{N-1}D_{\Phi}\left(A_{1}\cdots A_{k}:A_{k+1}\right).
\end{eqnarray}
Thus, we have
\begin{eqnarray}
D_{\Phi}(A_{1}:\cdots:A_{N})-D_{\Phi}(A_{1}:\cdots:A_{m})
=\sum_{k=m}^{N-1}D_{\Phi}\left(A_{1}\cdots A_{k}:A_{k+1}\right).
\end{eqnarray}
Similarly we can rewrite $D_{\Phi}(A_{1}:A_{m+1}:\cdots:A_{N})$ :
\begin{eqnarray}
D_{\Phi}(A_{1}:A_{m+1}:\cdots:A_{N})
=D_{\Phi}(A_{1}:A_{m+1})
+D_{\Phi}(A_{1}A_{m+1}:A_{m+2})
+\cdots+D_{\Phi}(A_{1}A_{m+1}\cdots A_{N-1}:A_{N}).\ \ \ \
\end{eqnarray}
Now we have
\begin{eqnarray}
&&D_{\Phi}(A_{1}:\cdots:A_{N})-D_{\Phi}(A_{1}:\cdots:A_{m})-D_{\Phi}(A_{1}:A_{m+1}:\cdots:A_{N})\nonumber\\
&=&\sum_{k=m}^{N-1}D_{\Phi}\left(A_{1}\cdots A_{k}:A_{k+1}\right)-[D_{\Phi}(A_{1}:A_{m+1})+D_{\Phi}(A_{1}A_{m+1}:A_{m+2})+\cdots+D_{\Phi}(A_{1}A_{m+1}\cdots A_{N-1}:A_{N})]\nonumber\\
&=&[D_{\Phi}\left(A_{1}\cdots A_{m}:A_{m+1}\right)-D_{\Phi}(A_{1}:A_{m+1})]+\cdots+[D_{\Phi}\left(A_{1} A_{m+1}:A_{m+2}\right)-D_{\Phi}(A_{1}A_{m+1}:A_{m+2})]+\cdots\nonumber\\
&+&[D_{\Phi}\left(A_{1}\cdots A_{N-1}:A_{N}\right)-D_{\Phi}(A_{1}A_{m+1}\cdots A_{N-1}:A_{N})].
\end{eqnarray}
We first minimize both sides of this equation with respect to $\Phi\left(\rho_{A_{1}\cdots A_{N}}\right)$.
Since we have that the bipartite GQDs do not increase under the discard of subsystems
and every item of the right hand side is greater than or equal to zero, it is obvious that
%\begin{eqnarray}
%D_{\Phi}(A_{1}:\cdots:A_{N})\geq D_{\Phi}(A_{1}:\cdots:A_{m})
%+D_{\Phi}(A_{1}:A_{m+1}:\cdots:A_{N}).
%\end{eqnarray}
\begin{eqnarray}
\underset{\Phi}{\min}D_{\Phi}(A_{1}:\cdots:A_{N})
&\geq&\underset{\Phi}{\min}D_{\Phi}(A_{1}:A_{2}:\cdots:A_{m})
+\underset{\Phi}{\min}D_{\Phi}(A_{1}:A_{m+1}:\cdots:A_{N})\\
\Rightarrow D(A_{1}:\cdots:A_{N})&\geq& D(A_{1}:A_{2}:\cdots:A_{m})
+D(A_{1}:A_{m+1}:\cdots:A_{N}).
\end{eqnarray}
Similarly, by using the condition that the bipartite GQDs do not increase
under the discard of subsystems, we can prove the general form of
these monogamy inequalities.
\end{proof}
\renewcommand{\theequation}{A2-\arabic{equation}}
% redefine the command that creates the equation no.
\setcounter{equation}{0}  % reset counter
\subsection*{A2.~Proof of Theorem 2}
\begin{proof}

In order to simplify our proof, we first consider a special case as
follows:
\begin{eqnarray}
D_{\Phi}(A_{1}:\cdots:A_{N})=D_{\Phi}(A_{1}:\cdots:A_{K})
+D_{\Phi}(A_{K+1}:\cdots:A_{N})+D_{\Phi}(A_{1}\cdots A_{K}:A_{K+1}\cdots A_{N}).
\end{eqnarray}
According to the definition of GQD, we have
$D(A_{1}:\cdots:A_{N})=\underset{\Phi}{\min}D_{\Phi}(A_{1}:\cdots:A_{N})$,
so we only need to consider the relationship between
$D_{\Phi}$. Using the definition, the $D_{\Phi}$ can be rewritten
as follows:
\begin{eqnarray}
D_{\Phi}(A_{1}:\cdots:A_{N})=I\left(\rho_{A_{1}\cdots A_{N}}\right)-I\left(\Phi\left(\rho_{A_{1}\cdots A_{N}}\right)\right).
\end{eqnarray}
Noting that the $I\left(\rho_{A_{1}\cdots A_{N}}\right)$ and $I\left(\Phi(\rho_{A_{1}\cdots A_{N}})\right)$ can be decomposed
as
\begin{eqnarray}
I\left(\rho_{A_{1}\cdots A_{N}}\right)&=&I\left(\rho_{A_{1}\cdots A_{K}}\right)+I\left(\rho_{A_{K+1}\cdots A_{N}}\right)
+I\left(\rho_{\left(A_{1}\cdots A_{K}\right)\left(A_{K+1}\cdots A_{N}\right)}\right)\\
I\left(\Phi(\rho_{A_{1}\cdots A_{N}})\right)&=&I\left(\Phi(\rho_{A_{1}\cdots A_{K}})\right)+I\left(\Phi(\rho_{A_{K+1}\cdots A_{N}})\right)
+I\left(\Phi(\rho_{\left(A_{1}\cdots A_{K}\right)\left(A_{K+1}\cdots A_{N}\right)})\right).
\end{eqnarray}
%Then we have
%\begin{eqnarray}
%D_{\Phi}(A_{1}:\cdots:A_{N})
%=[I\left(\rho_{A_{1}\cdots A_{K}}\right)+I\left(\rho_{A_{K+1}\cdots A_{N}}\right)+I\left(\rho_{\left(A_{1}\cdots A_{K}\right)\left(A_{K+1}\cdots A_{N}\right)}\right)]-[I\left(\Phi\left(\rho_{A_{1}\cdots A_{K}}\right)\right)+I\left(\Phi\left(\rho_{A_{K+1}\cdots A_{N}}\right)\right)+I\left(\Phi\left(\rho_{\left(A_{1}\cdots A_{K}\right)\left(A_{K+1}\cdots A_{N}\right)}\right)\right)]\nonumber\\
%=[I\left(\rho_{A_{1}\cdots A_{K}}\right)-I\left(\Phi\left(\rho_{A_{1}\cdots A_{K}}\right)\right)]+[I\left(\rho_{A_{K+1}\cdots A_{N}}\right)-I\left(\Phi\left(\rho_{A_{K+1}\cdots A_{N}}\right)\right)]+[I\left(\rho_{\left(A_{1}\cdots A_{K}\right)\left(A_{K+1}\cdots A_{N}\right)}\right)-I\left(\Phi\left(\rho_{\left(A_{1}\cdots A_{K}\right)\left(A_{K+1}\cdots A_{N}\right)}\right)\right)]\nonumber\\
%=D_{\Phi}(A_{1}:\cdots:A_{K})
%+D_{\Phi}(A_{K+1}:\cdots:A_{N})+D_{\Phi}((A_{1}\cdots A_{K}):(A_{K+1}\cdots A_{N}))
%\end{eqnarray}
Thus, this particular case has been proved.

Furthermore, we can prove the general form of this identity by using
the similar decomposition of $I\left(\rho_{A_{1}\cdots A_{N}}\right)$
as follows:
\begin{eqnarray}
I\left(\rho_{A_{1}\cdots A_{N}}\right)=I\left(\rho_{A_{1}\cdots A_{K_{1}}}\right)+\cdots+I\left(\rho_{A_{K_{N}+1}\cdots A_{N}}\right)
+I\left(\rho_{\left(A_{1}\cdots A_{K_{1}}\right)\cdots\left(A_{K_{N}+1}\cdots A_{N}\right)}\right),
\end{eqnarray}
which completes the proof.
\end{proof}
\renewcommand{\theequation}{A3-\arabic{equation}}
% redefine the command that creates the equation no.
\setcounter{equation}{0}  % reset counter
\subsection*{A3.~Proof of Proposition 1}
\begin{proof}
In order to prove the above inequality, first of all, we use $D_{T}$,
$D_{K_{1}}$, $D_{K_{2}}$, $D_{K_{3}}$, $\cdots$ , $D_{N}$ and
$D_{I}$ to represent $D(A_{1}:\cdots:A_{N})$, $D(A_{1}:\cdots:A_{K_{1}})$, $D(A_{K_{1}+1}:\cdots:A_{K_{2}})$,
$D(A_{K_{2}+1}:\cdots:A_{K_{3}})$, $\cdots$, $D(A_{K_{N}+1}:\cdots:A_{N})$, $D(A_{1}\cdots A_{K_{1}}:A_{K_{1}+1}\cdots A_{K_{2}}:A_{K_{2}+1}\cdots A_{K_{3}}:\cdots:A_{K_{N}+1}\cdots A_{N})$. Now we need to show that

\begin{eqnarray}
[D_{T}]^{n}\geq[D_{K_{1}}]^{n}+[D_{K_{2}}]^{n}+[D_{K_{3}}]^{n}+\cdots+[D_{N}]^{n}+[D_{I}]^{n}.
\end{eqnarray}

According to the above Theorem~2, we have
\begin{eqnarray}
[D_{T}]^{n}\geq%[D_{K_{1}}+D_{K_{2}}+D_{K_{3}}+\cdots+D_{N}+D_{I}]^{n}
[(D_{K_{1}}+D_{K_{2}}+D_{K_{3}}+\cdots+D_{N})+D_{I}]^{n}.
\end{eqnarray}
Using the binomial theorem, the above formula can be re-expressed
as
\begin{eqnarray}
[D_{T}]^{n}\geq\sum_{k=0}^{n}C_{n}^{k}(D_{K_{1}}+D_{K_{2}}+D_{K_{3}}+\cdots+D_{N})^{n-k}D_{I}^{k}
=(D_{K_{1}}+D_{K_{2}}+D_{K_{3}}+\cdots+D_{N})^{n}+D_{I}^{n}
\notag\\
+\sum_{k=1}^{n-1}C_{n}^{k}(D_{K_{1}}+D_{K_{2}}+D_{K_{3}}+\cdots+D_{N})^{n-k}D_{I}^{k}
\geq(D_{K_{1}}+D_{K_{2}}+D_{K_{3}}+\cdots+D_{N})^{n}+D_{I}^{n}.
\end{eqnarray}
Similarly, by using the same procedure many times, we have
\begin{eqnarray}
[D_{T}]^{n}\geq(D_{K_{1}}+D_{K_{2}}+D_{K_{3}}+\cdots+D_{N})^{n}+D_{I}^{n}
\geq(D_{K_{1}}+D_{K_{2}}+D_{K_{3}}+\cdots+D_{K_{N}})^{n}+D_{N}^{n}+D_{I}^{n}
\notag\\
\cdots\ \ \ \ \ \ \ \ \ \ \ \ \ \ \ \ \ \ \
\notag\\
\geq(D_{K_{1}}+D_{K_{2}})^{n}+D_{K_{3}}^{n}+\cdots+D_{N}^{n}+D_{I}^{n}
\geq[D_{K_{1}}]^{n}+[D_{K_{2}}]^{n}+[D_{K_{3}}]^{n}+\cdots+[D_{N}]^{n}+[D_{I}]^{n},
\end{eqnarray}
which completes the proof.
\end{proof}

\renewcommand{\theequation}{A4-\arabic{equation}}
% redefine the command that creates the equation no.
\setcounter{equation}{0}  % reset counter
\subsection*{A4.~Proof of Theorem 3}
\begin{proof}

In order to prove the above inequality, we only need to consider the
relationship between
$D_{\Phi}$. By using property $\left(a\right)$,
$D_{\Phi}\left(A_{1}:\cdots:A_{N}\right)$ can be decomposed
as follows:
\begin{eqnarray}
D_{\Phi}(A_{1}:\cdots:A_{N})&=&D_{\Phi}(A_{1}:A_{2}:\cdots:A_{K})+D_{\Phi}(A_{1}\cdots A_{K}:A_{K+1})+D_{\Phi}(A_{1}\cdots A_{K+1}:A_{K+2})+\cdots+D_{\Phi}(A_{1}\cdots A_{N-1}:A_{N})\nonumber\\
&=&D_{\Phi}(A_{1}:A_{2}:\cdots:A_{K})+\sum_{j=K}^{N-1}D_{\Phi}(A_{1}\cdots A_{j}:A_{j+1}).\ \ \ \ \ \ \ \
\end{eqnarray}
Then we have
\begin{eqnarray}
D_{\Phi}(A_{1}:\cdots:A_{N})-\sum_{i=1}^{N-K}D_{\Phi}(A_{i}\cdots A_{i+K-1}:A_{i+K})
&=&D_{\Phi}(A_{1}:A_{2}:\cdots:A_{K})
+[D_{\Phi}(A_{1}\cdots A_{K+1}:A_{K+2})-D_{\Phi}(A_{2}\cdots A_{K+1}:A_{K+2})]\nonumber\\
&+&\cdots+[D_{\Phi}(A_{1}\cdots A_{N-1}:A_{N})-D_{\Phi}(A_{N-K}\cdots A_{N-1}:A_{N})].
\end{eqnarray}
After minimizing both sides of this equation with respect to $\Phi\left(\rho_{A_{1}\cdots A_{N}}\right)$,
we have that the bipartite GQDs do not increase under the discard
of subsystems and every item of the right hand side is greater than or
equal to zero.  We therefore have
%\begin{eqnarray}
%D_{\Phi}(A_{1}:\cdots:A_{N})-\sum_{i=1}^{N-K}D_{\Phi}(A_{i}\cdots A_{i+K-1}:A_{i+K})
%\geq D_{\Phi}(A_{1}:A_{2}:\cdots:A_{K}).
%\end{eqnarray}
\begin{eqnarray}
D(A_{1}:\cdots:A_{N})-\sum_{i=1}^{N-K}D(A_{i}\cdots A_{i+K-1}:A_{i+K})
\geq D(A_{1}:A_{2}:\cdots:A_{K})\geq0,
\end{eqnarray}
which completes the proof.
\end{proof}

\end{widetext}

\end{document}